\documentclass[a4paper]{jpconf}
\usepackage{graphicx}
\begin{document}
\begin{flushleft} 
KCL-PH-TH/2011-34 \\
LCTS/2011-17 \\
\end{flushleft} 

\title{Neutrinos and the Universe}

\author{Nick E. Mavromatos}

\address{Theoretical Particle Physics and Cosmology Group, Department of Physics, King's College London,  Strand, London WC2R 2LS, UK , and \\Theory Division, Physics Department, CERN, CH 1211 Geneva 23, Switzerland}

\ead{Nikolaos.Mavromatos@kcl.ac.uk}

\begin{abstract}
In this talk, I review the potential connection of neutrinos to the Physics of the Early Universe, in particular the r\^ole of (sterile) neutrinos to leptogenesis/baryogenesis and the dark sector of the Universe. The possibility of CPT Violation among active neutrinos at early times  and its role in leptogenesis/baryogenesis without sterile neutrinos is also touched upon.

 \end{abstract}
\begin{center} 
\vspace{-1cm}
Invited paper to
NUFACT 11, \emph{XIIIth International Workshop on Neutrino Factories, Super beams and Beta beams}, 1-6 August 2011, CERN and University of Geneva \\
(Submitted to IOP conference series)
\end{center} 

\section{Introduction and Motivation}
\paragraph{} 
Neutrinos are fascinating particles, still full of mysteries and surprises. Not long  after the end of this meeting, we have heard the claims from OPERA Collaboration~\cite{opera} on the measurements of the arrival time of neutrinos, indicating \emph{superluminal} propagation. Although this result may not be actually due to fundamental  (Lorentz violating or modifying) physics but rather due to measurement uncertainties, given the complicated nature of the measurement and the many potential systematic errors that may have entered (including the particle decays that can produce the final signal, which undoubtedly introduce statistical uncertainties in the arrival times), nevertheless neutrinos made big headlines all over the world once again, and opened up interesting avenues for research, given that superluminal propagation, that characterises already existing theoretical models, may not be incompatible with causality. 

In this talk I will not discuss such issues, however I will touch upon a different r\^ole of neutrinos (assumed throughout to respect Lorentz Invariant kinematics, thus being \emph{subluminal}, almost light-like due to their tiny masses) that relates to the Physics of the Early Universe. In particular, one of 
the most important questions of fundamental Physics that is still unanswered today, which pertains to our very existence, is the reason for the observed baryon asymmetry in the Universe, or why the Universe is made up mostly of matter. According to the Big bang theory, matter and antimatter have been created at equal amounts in the Early universe. The observed charge-parity (CP) violation in particle physics~\cite{cronin}, prompted A. Sakharov~\cite{sakharov} to conjecture that non-equilibrium Physics in the Early Universe produce Baryon number (B), charge (C) and charge-parity (CP) violating, but CPT \emph{conserving},  interactions/decays of anti-particles in the early universe, resulting in the observed baryon-antibaryon ($n_B - n_{\overline{B}}$) asymmetry. 
In fact there are two types of non-equilibrium processes in the Early universe that could produce this asymmetry: the first type concerns processes generating asymmetries between leptons and antileptons, while the second produce asymmetries between baryons and antibaryons. 

The almost 100\% observed asymmetry today, is estimated in the Big-Bang theory~\cite{gamow} to be of order: 
\begin{equation}\label{basym}
\Delta n (T \sim 1~{\rm GeV}) = \frac{n_{B} - n_{\overline{B}}}{n_{B} + n_{\overline{B}}} \sim \frac{n_{B} - n_{\overline{B}}}{s} = (8.4 - 8.9) \times 10^{-11} 
\end{equation}
at the early stages of the expansion, e.g. for times $t < 10^{-6}$~s and 
temperatures $T > 1 $~GeV. In the above formula $n_B$ ($n_{\overline{B}}$) denotes the (anti) baryon density in the Universe, and $s$ is the entropy density. 

Unfortunately, the observed CP violation within the Standard Model (SM) of Particle Physics (found to be of order $\epsilon = O(10^{-3})$  in the neutral Kaon experiments~\cite{cronin}) cannot reproduce (\ref{basym}). Let us review why~\cite{kuzmin,shapo}.
  
Although classically, the baryon (B) and Lepton (L$_{e,\mu,\tau}$) numbers are conserved in the SM, at a  quantum level  \emph{anomalies}  in general break these symmetries. The anomalies are associated with the non-conserved currents corresponding to the above  classical symmetries in the combination
$ \partial^\mu  J_\mu^{B} = \partial^\mu J_\mu^{L}  = \frac{N_f}{32\pi^2} {\rm Tr}F_{\mu\nu} \, \tilde{F}^{\mu\nu} + {\rm U}(1)-{\rm parts} $, with 
$N_f$ the flavour number. Since the allowed processes in the SM are those which entail a change of B by multiples of 3, that is bosons 
$\leftarrow\rightarrow$ bosons + 9 quarks + 3 leptons, there is a conservation law for the three combinations $L_i - B/3$, $i=e,\mu,\tau$.
However the observed neutrino oscillations among flavours, imply that only one global number may be conserved in the SM, the combination $B-L$,
where $L = \sum_{i} L_i$ is the total Lepton number. In fact if neutrinos are Majorana, the L would be itself violated, and hence there would be no conserved numbers at all!

With this in mind one may evaluate the rate of B violation in the SM~\cite{kuzmin,shapo}: $\Gamma \sim (\alpha_W T)^4\, (M_{\rm sph}/T)^7 {\rm exp}(-\frac{M_{\rm sph}}{T})$, if $T \le M_{\rm sph}$, while $\Gamma \sim (\alpha_W T)^4 \alpha_W {\rm log}(1/\alpha_W) $ if $T \ge M_{\rm sph}$, with 
$\alpha_W$ the fine structure constant of the electroweak SU(2) symmetry, and $M_{\rm sph}$ is the spaleron mass scale, in particular 
$M_{\rm sph}/\alpha_W$  is the height of the energy barrier separating SU(2) vacua with different topologies. 
Thermal equilibrium (i.e. $\Gamma > H$ (Hubble rate)) for B non conserving processes occurs only for~\cite{kuzmin}
$T_{\rm sph} (m_h) < T < \alpha_W^5 M_{\rm P} \sim 10^{12}$~GeV;  with the Higgs mass $m_h$ assumed in the range (in agreement with current LHC exclusion data) $m_h \in [100, 300]$ GeV , one has $T_{\rm sph} \in [130, 190]$~GeV. Baryon Asymmetry in the Universe (BAU) could be produced only when the sphaleron interactions freeze out, that is for temperatures $T \simeq T_{\rm sph}$. One should compute, within the SM, the CP violation effects at such a regime of parameters and temperature, using the Cabbibo-Kobayashi-Maskawa (CKM) matrix. The lowest SM CP violating structures are encoded in the quantity $\delta_{CP} = {\rm sin}(\theta_{12}) {\rm sin}(\theta_{23}) {\rm sin}(\theta_{13}) {\rm sin}\delta_{CP} (m_t^2 - m_c^2) 
(m_t^2 - m_u^2) (m_c^2 - m_u^2) (m_b^2 - m_s^2) (m_b^2 - m_d^2) (m_s^2 - m_d^2)$, where $\delta_{CP} = D/T^{12}$  is the Kobayashi-Maskawa CP Violating phase, and $D$ is the Jarlskog determinant, related to the appropriate quark mass matrices as~\cite{shapo}: $D= {\rm Tr}(\mathcal{M}_u^2 \mathcal{M}_d^2 \mathcal{M}_u \mathcal{M}_d)$. Computing the parameter $\delta_{CP}$ at $T \simeq T_{\rm sph}$ in the above range, 
one obtains $\delta_{CP} \sim 10^{-20}$ which yields a baryon asymmetry ten orders of magnitude smaller than the observed one (\ref{basym}). 
Hence the SM CP violation cannot be the source for the observed BAU. 

There are several ideas that go beyond the SM (e.g. GUT models, Supersymmetry, extra dimensional models etc.)
in an attempt to find extra sources for CP violation that could generate the observed BAU. 
Since massive neutrinos (as evidenced by the observed flavour oscillations) constitute the simplest extension of SM, it is reasonable to 
seek for a possible r\^ole of neutrinos in providing us with the required amount of CP violation to explain the observed BAU. 
Right-handed supermassive (Majorana) neutrinos (\emph{sterile)} may do the job and more than that. Namely, as we shall discuss below,  
such extensions of the SM can provide  sufficient  extra sources for CP Violation to explain the Origin of Universe's matter-antimatter asymmetry due to the relevant neutrino masses, without the need for Supersymmetry or extra dimensions. Moreover, as we shall discuss below, such models can also incorporate a natural  Dark Matter candidate (the lightest of the sterile neutrinos), in agreement with observations~\cite{shapo}. 

The structure of the remainder of the talk is the following: in the next section \ref{sec:2}, I discuss the simplest (non supersymmetric) extensions of the SM involving sterile neutrinos and their connection with leptogenesis/baryogenesis. In section \ref{sec:3},  I discuss the connection of neutrinos to the Dark Sector of the Universe, including the r\^ole of light sterile neutrinos as Dark Matter candidate, as well as exotic possibilities such as neutrino condensates as contributions to dark energy, and mass varying neutrinos and their contribution to the accelerating cosmic expansion. In section \ref{sec:4}, I discuss CPT Violation in the early Universe as a possible way to avoid sterile neutrinos, but still involving light (active) neutrino-antineutrino asymmetries to generate the observed BAU. Due to lack of available space my discussions will be very brief, referring the interested reader for details to the rather vast literature on  the subject, through appropriate reviews and references that I used. I apologise beforehand for omissions in citations, but this is the best I can do given the length restrictions. 

\section{Sterile Neutrino SM Extensions and Leptogenesis/Baryogenesis \label{sec:2}}
\paragraph{} 

Several authors have suggested the use of right-handed, supermassive sterile neutrinos as possible extensions beyond the SM, with relevance to the physics of the Early universe. In this talk, I concentrate on the simplest of such extensions, a non supersymmetric SM augmented with N generations of right-handed massive, with masses $M_I$, Majorana fermions, termed $\nu MSM$, a terminology I will use from now on:~\cite{shapo,nsm}
\begin{equation}\label{nMSM}
L_{\nu {\rm MSM}} =   L_{{\rm SM}} + {\overline N}_I i\gamma^\mu \partial_\mu N_I -F_{\alpha I} {\overline L}_\alpha N_I {\tilde \phi} -\frac{M_I}{2} {\overline N}^c_I N_I + {\rm h.c} 
\end{equation}
where the suffix SM denotes the SM part of the Lagrangian, $N_I$, $I=1, \dots N$ denote the Majorana sterile neutrinos, the superscript $c$ denotes charge conjugate, and $L_\alpha$, $\alpha = e,\mu,\tau$ are the leptons. The field $\tilde{\phi}$  is the SU(2) dual of the Higgs scalar $\tilde \phi = \epsilon_{ij} \phi^\star_j $,
$i,j$ SU(2) indices, while $F_{\alpha I}$ are matrix valued Yukawa couplings involving majorana phases and mixing angles~\cite{shapo,nsm}. 
The model with N=1 sterile neutrino is excluded by the current data~\cite{giunti}, while the models with N=2,3 work well in reproducing BAU and are consistent with the current experimental data on neutrino oscillations. Of particular interest to us will be the Model with N=3 singlet neutrinos,  which in fact allows one of the Majorana fermions  to almost decouple from the rest of the SM fields, thus providing 
candidates for light (KeV region of mass) sterile neutrino Dark Matter~\cite{nsm}.  Moreover, these models can also be consistent with inflationary scenarios, through, e.g., non-minimal couplings of the Higgs scalars to the Einstein curvature tensor~\cite{berz}, 
$L_G \in \zeta \phi^\dagger \phi R + \dots $, with flat effective potential
for large values of $\phi$, thus being compatible with the inflation slow roll conditions. We shall not discuss such important issues here though. 
The light (active) neutrino $\nu$ masses in the model are generated through appropriate see-saw mechanisms, $m_\nu = -M^D \frac{1}{M_I} [M^D]^T$,
with T indicating matrix transposition, $M^D = F_{\alpha I} v$, $v = <\phi > $ is the standard model Higgs v.e.v., assumed in \cite{nsm} to be of order 175 GeV thereby yielding  $M_D \ll M_I$. 

For the connection with Leptogenesis/Baryogenesis we need to know the thermal properties of the model. A detailed analysis~\cite{nsm,ars}, which we shall omit here, indicates that the relevant decay processes in the early Universe, which -- when out of equilibrium could lead to matter-antimatter asymmetries, that is $N t \leftarrow\rightarrow \nu\, h ~, \, h \leftarrow\rightarrow N \nu, \, N \leftarrow\rightarrow h\, \nu$, where $h$ is the Higgs field, 
have a rate $\Gamma \sim 9 F^2 f_t^2 T/(64\pi^3)$, whereby $f_t$ denotes the top quark Yukawa coupling, and the amplitude of the sterile Neutrinos Yukawa couplings $|F|^2 \sim v^{-2} \, m_{\rm atm} M_I \sim 2 \times 10^{-15} \frac{M_I}{{\rm GeV}}$ , $m_{\rm atm}^2 \equiv |\Delta m_{\rm atm} |^2 
= (2.40 + 0.12 - 0.11) \times 10^{-3}$~eV$^2$, is the measured neutrino mass difference form atmospheric experiments. From this rate the conditions for thermal equilibrium (i.e. $\Gamma > H$(Hubble)) can be computed and the equilibrium temperature  is $T_{\rm eq} \simeq \frac{9\, f_t^2 
m_{\rm atm} M_0}{64\pi^3 \, v^2} M_I \simeq 5 M_I, $ with $M_0 = M_P/(1.66 \sqrt{g_{\rm eff}})$, with $M_P$ the (four dimensional) Planck mass and $g_{\rm eff}$ the effective degrees of freedom in the radiation era of the Universe. There are two physically distinct cases we consider: $M_I \ge M_W $ and $M_I < M_W$, where $M_W$ the electroweak breaking scale ($M_W \sim $ 100 GeV).

\emph{(I) Case where $M_I \ge M_W$} : In the first case, the decay of the right-handed fermions occurs at tempertures~\cite{nsm,shapo} $T_{\rm decay} = \left(\frac{m_{\rm atm} M_0 }{24\pi^2 \, v^2 }\right)^{1/3}\, M_I \simeq 3M_I$. Such processes are out of equilibrium for $T > T_{\rm eq} \sim 5M_I $ or $T < T_{\rm decay} \sim 3M_I$. If $T_{\rm eq} > 
T_{\rm sph} $, then the decays of the right-handed fermions occur during the period for which the sphaleron processes are active, i.e. $T_{\rm decay} > T_{\rm sph}$, this leads to thermal leptogenesis~\cite{lepton} as follows:  

The Heavy Right-handed Majorana neutrinos enter equilibrium at $T = T_{\rm eq}  > T_{\rm decay}$, which in fact is independent of the initial conditions at $T \gg T_{\rm eq}$. Subsequently, as the Universe continues to expand and cools down, 
the out of equilibrium \emph{ Lepton Number violating} decays of the heavy neutrinos at $T \simeq T_{\rm decay} > T_{\rm sph}$: $N _I \rightarrow \nu \, h~, \,  {\overline \nu} \, {\overline h} $ produce a Lepton-antilepton \emph{asymmetry }, which is then communicated (through the induced  effective low-energy Lepton - number violating interactions in the SM Lagrangian of the form $L \ni \frac{2}{M_I}{\overline L}_L L_L \phi \phi + {\rm h.c.} $, where $L_L$ denote the SM left-handed lepton doublets)
to the baryon sector through 
equilibrated B+L violating sphaleron interactions, independent of the initial conditions, to produce the observed Baryon Asymmetry.

\begin{figure}[h]
\begin{center}
\includegraphics[width=0.5\textwidth]{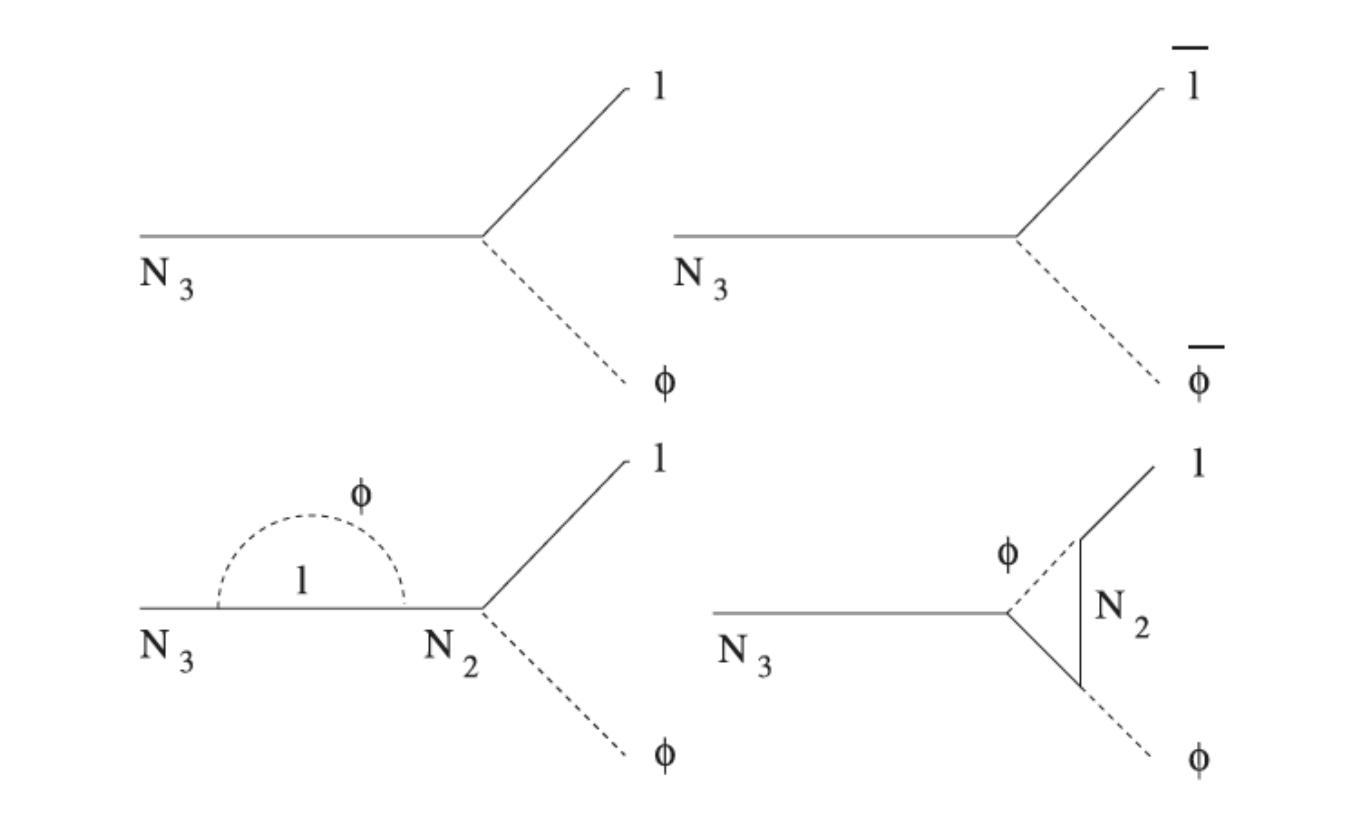} \hfill \includegraphics[width=0.4\textwidth]{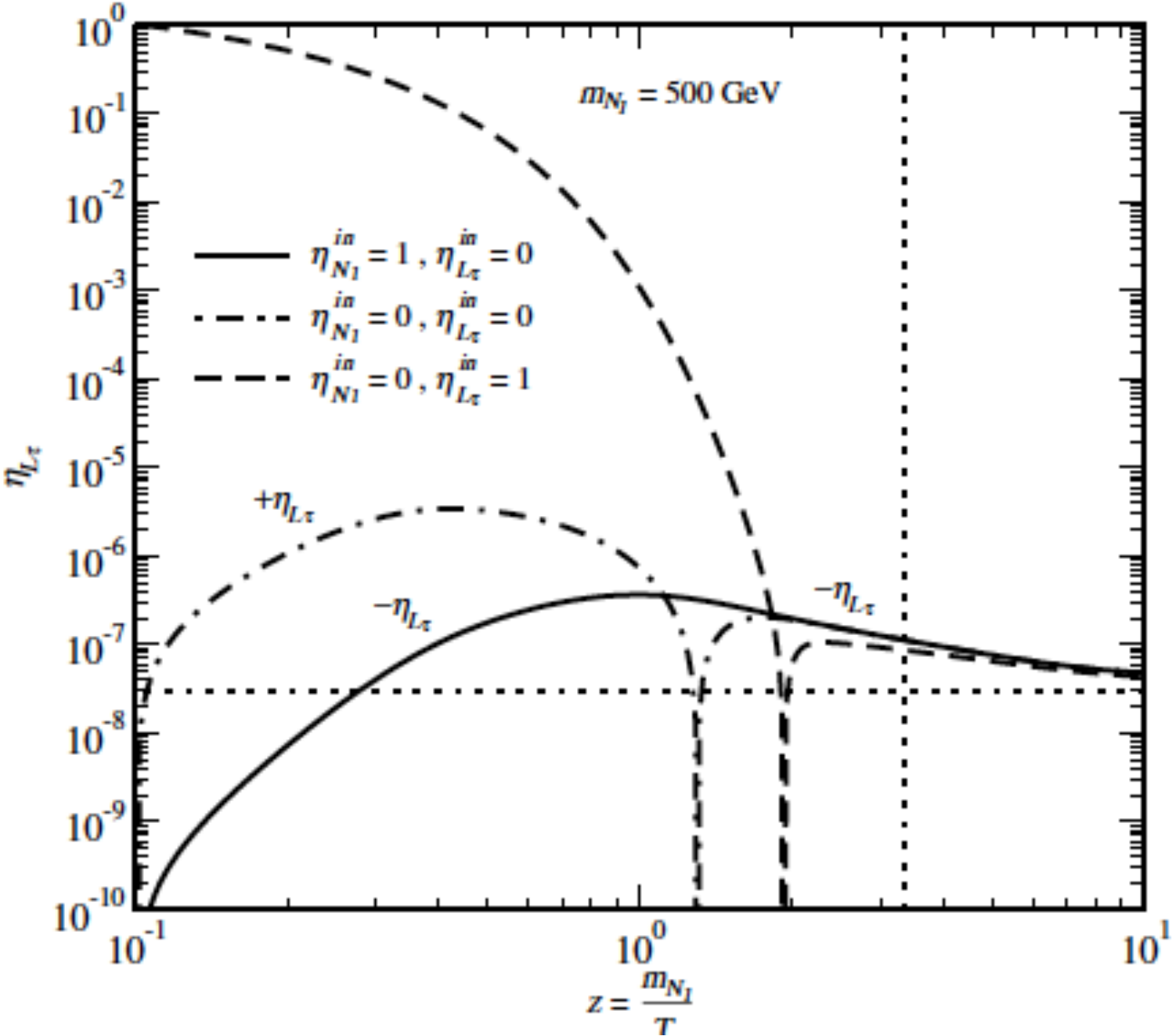} \end{center}
\caption{\label{fig:boltz} \emph{Left picture:} relevant graphs contributing to the BAU in the $\nu$MSM model with three sterile neutrinos. The lower two graphs are responsible for enhancement of CP violation if the two heavy sterile neutrinos $N_{2,3}$ are degenerate in mass. \emph{Right picture:} An example of heavy $N_1$ neutrino  abundance: thermal relic density $\eta_{N_1}$ vs $m_{N_1}/T$ in the scenario  of \cite{pilaftsis} of resonant Leptogenesis. }
\end{figure}
At this point we make some important remarks: to calculate the precise contributions to the BAU and compare it with the observed one, one needs to 
estimate the (thermal) relic abundance $\eta_{N_i}$ of the heavy neutrinos by solving the appropriate Boltzmann equations, 
$\frac{d}{d t} \eta_{N_i} + 3 H \eta_{N_i} = C[f]$, $\eta_{N_i} \equiv \int d^3 p f_N(p, x, t) $. The results are sensitive to the underlying theoretical model used, of course~\footnote{It must be noted that the Boltzmann equations are classical
phase space equations, giving the 
time evolution of phase space
distribution functions f(p,x.t); however, the Collision terms $C[f]$ on the 
r.h.s. are quantum effects involving loop corrected 
cross sections (particularly in Leptogenesis scenarios). Since the Leptogenesis processes are out of (thermal) equilibrum, 
the correct approach to the problem, incorporating properly the quantum effects, would be to consider non-equilibrium field theoretic methods
to calculate directly Heavy neutrino 
abundances by using 
the non-equlibrium field theory
Schwinger-Keldysh formalism and the 
Kadanoff-Baym (KB) equations,  
based on Green's functions, not on densities. 
Such a programme has been undertaken recently in \cite{buchmuller}, with the conclusion 
that conventional Boltzmann
equations for Lepton asymmetry give pretty good agreement with 
quantum field.theory treatments
when SM gauge interactions are
taken into account.} (cf. fig.~\ref{fig:boltz}). If in the $\nu$MSM model, we concentrate here for definiteness, with N=3 heavy majorana neutrinos,
 all neutrino flavours are non degenerate in mass, $| M_I - M_J | \sim M_K$, then unfortunately, in order  to reproduce the correct BAU  (with the ratio
 $n_B/s \sim F^2 10^{-3} \sim 10^{-10}$ (\emph{cf.} (\ref{basym}))) one needs Yukawa couplings of order $F^2 \sim 10^{-7}$~\cite{shapo,nsm}, which yield masses for the heavy neutrinos of order $M_{N_I} = 10^{11}$ GeV. Such a scenario is plagued by instabilities of the Higgs mass under quantum corrections, e.g. the one-loop corrections are found to be of order $F^2M_I^2 / (4\pi) \sim 10^{14} {\rm GeV}^2$.
A resolution to this problem is provided by considering models in which two of the heavy neutrinos are degenerate in mass, say $N_{2,3}$~\cite{pilaftsis,riotto,nsm,shapo}. In such a case, much smaller Yukawa couplings are allowed, $f^2 \sim \frac{M_2 - M_3}{M_2} \sim \frac{m_\nu \, M_W}{v^2} \sim 10^{-13}$, with $M_I \sim M_W$. There is an enhancement of the induced CP violation in this case (due to the lower two graphs in the left picture of fig.~\ref{fig:boltz}). This enhanced CP violation due to the existence of degenerate in mass quantum states is familiar from the case of neutral Kaons~\cite{cronin}. 

Such models yield $n_B/s \sim 10^{-3} f^2 \frac{M_2 \Gamma_{\rm tot}}{(M_2 - M_3)^2 + 
\Gamma_{\rm tot}^2} $, with $|M_2 - M_3 | \sim \Gamma_{\rm tot}$, and $\Gamma_{\rm tot}$ the total sterile neutrinos decay width.
For $M_I < 10^7 $ GeV there is no problem with Higgs stability in such models~\cite{shapo}. 
A specific example has been considered in \cite{pilaftsis} in which only one of the heavy neutrinos, $N_1$, decays out of equilibrium, while the heavier ones $N_{2,3}$ stay in thermal equilibrium.  One lepton number ($\tau$)  is resonantly produced by the out-of-equilibrium decays of $N_1$. 
To avoid excess of L$_\tau$ - number violation, the decay rates of $N_{2,3}$ are suppressed. 
The predicted BAU in such a class of models (termed resonant Leptogenesis for obvious reasons) can be naturally made to agree with the observed one, as following from a calculation of the thermal relic abundances of the $N_1$ neutrino, by means of solving the pertinent Boltzmann equations (cf. right picture in fig.~\ref{fig:boltz}). Leptogenesis is possible for these models if $M_I $ is in the range $M_I \in [M_W - {\rm TeV}]$. The predicted Lepton-number violating process $\mu \to e\, \gamma$ has branching ratio of order $B(\mu \to e \gamma) = 6 \times 10^{-4} a^2 b^2 v^4/M_I^4 $, where $a, b$ are parameters entering the appropriate Yukawa coupling matrix of the model~\cite{pilaftsis}. For natural values $a, b \sim 10^{-3}$, required by Leptogenesis, one has $B(\mu \to e \gamma) \le 1.2. \times 10^{-11}$ for the above range of $M_I$. This is just one order of magnitude larger than the 
current sensitivity of the MEG Experiment~\cite{meg} to the Branching ratios for $\mu^+ \to e^+ \, \gamma$, $B(\mu^+ \to e^+ + \gamma )  \le 2.4 \times 10^{-12}$. In ref. \cite{pilaftsis} there were also examined possible effects of such degenerate models in linear $e^+ e^-$ colliders, where one should study the production of
electroweak scale $N_{2,3}$ via their decays to $e, \, \mu$ but not $\tau$. 

\emph{(I) Case where $M_I < M_W$} : consider, for instance, the case where $M_I =$ O(1) GeV. In such a case, in order to guarantee the smallness of the light neutrino masses, the relevant Yukawa couplings should be of order~\cite{shapo}:  $F_{\alpha I} \sim \frac{\sqrt{m_{\rm atm} M_I}}{v} \sim  
4 \times 10^{-8}$. One may further assume that two of the heavy majorana neutrinos are degenerate in mass, say $N_{2,3}$ as before, which enhances CP violation. 
In such a case, the scenario of ref. \cite{ars}, on Baryogenesis through coherent oscillations of the two degenerate in mass right-handed fermions, 
may be realised. The Heavy Majorana fermions $N_I$ 
thermalize only for $T < M_W$, so their decays at for $T > M_W$  are out of equilibrium. One may worry in such a case that the induced  BAU
would depend on the initial conditions. However, inflation may set the initial concentration of $N_I$ to practically zero value at the end of inflation.
The relevant Majorana masses are small compared to the Sphaleron freeze-out Temperature, hence the total lepton number is conserved (the total Lepton number is zero but unevenly distributed between active and sterile neutrinos), this leads to ``apparent'' lepton number violation, so that 
the Lepton number of active left-handed $\nu$ is  transferred to Baryons due to 
equilibrated sphaleron processes. The coherent oscillations between the mass-degenerate singlet fermions have a frequency~\cite{ars,shapo}:
$\omega \sim |M_2^2 - M_3^2|/E_I \sim M_2 \Delta M /T$, for $\Delta M \equiv |M_2 - M_3| \ll M_2 \simeq M_3, E_I \sim T$.
For CP violation to occur (so that one can achieve the (observed) BAU) one must have that the oscillation rate is larger than the Hubble expansion rate of the Universe. Under the (important and rather delicate) assumption that the interactions with the plasma of SM particles in the early universe 
do not destroy quantum mechanical coherence of oscillations, one may have in this model baryogenesis occurring at 100 GeV, and a maximal baryon asymmetry $\Delta = n_B - n_{\overline B}/( n_B +  n_{\overline B}) \sim 1$ when $T_B \simeq T_{\rm sph} \simeq  T_{\rm eq}$. Thus this mechanism for producing BAU seems quite effective, if in operation. For the case $T_B \gg T_{\rm sph} > T_{\rm eq}$ the predicted $n_B/s$ in this version of the $\nu$MSM model reads~\cite{shapo}: $n_B/s \sim 1.7 \times 10^{-10} \delta_{CP} \, 
\left(\frac{10^{-5}M_2}{\Delta M(T)}\right)^{2/3} \, \left(\frac{M_2}{10~{\rm GeV}}\right)^{5/3} $, with the CP violating factor expressed through the various mixing angles and CP-violating phases. This can be of O(1), according to the recent experimental data of neutrino oscillations. The observable value of the BAU, then, can be obtained for a wide range of parameters of the $\nu$MSM~\cite{shapo}. In particular, one may have 
Mass $N_2$ ($N_3$) / (Mass $N_1$)  = O($10^5$ ), implying that the lightest sterile neutrino may have masses in the keV range,  if $M_{2,3}$ are of O(1-10) GeV. As we shall argue in the next section, this Lightest Sterile neutrino of this version of $N=3$ $\nu$MSM is a 
natural Dark Matter (DM) candidate~\cite{shapo,nsm}. 

Before closing this section we should mention that there are several theoretical scenarios that can explain the required mass hierarchy $M_{2,3} \gg M_{1}$ among the sterile neutrinos. These scenarios range from: (i)  the imposition of flavour symmetries~\cite{shapo,mohapatra,merle} (one starts with $M_1 = 0$ and $M_2 \simeq M_3 > {\rm GeV}$ if symmetry is unbroken, while Breaking of global Lepton symmetry generate singlet fermion mass hierarchy, $M_1 =O({\rm keV}) \ll M_{2,3}$), to (ii)  brane world scenarios~\cite{kushenko} exploiting the associated 
exponential factor in a Randall-Sundrum-like framework~\cite{rs} to obtain a large mass splitting  (one sterile neutrino at the keV scale,
while the other two could have masses of around $10^{11}$ GeV or  heavier, and  at the
same time ensuring that the seesaw mechanism for active neutrinos works) from very moderately tuned parameters, and (iii) to Froggatt-Nielsen type mechanisms~\cite{merle2}, whereby one fermion  acquires mass via Higgs mechanism, while the rest via higher order multiple see-saw.

\section{Neutrinos and the Dark Sector of the Universe \label{sec:3}}

\emph{Sterile neutrinos in $\nu$MSM and DM}: To be DM, the lightest massive sterile neutrino of the $N=3$ $\nu$MSM, $N_1$, must have a life time larger than that of the Universe. For this to happen, its coupling with SM matter must be superweak, 
$\theta_1 G_F = \sum_{\alpha =e,\mu,\tau}\,\frac{v^2 |F_{\alpha I}|^2 }{M_1^2}$, with $G_F$ denoting the Fermi coupling of the weak interactions.
The width of $N_1$ for the decay $N_1 \rightarrow \gamma \, \nu$ is expressed in terms of $\theta_1$ mixing angle as follows~\cite{shapo}:
\begin{equation}\label{widthn1}
\Gamma_{N_1 \to \gamma \nu} = \frac{9 \alpha G_F^2}{1024\pi^4}{\rm sin}^2 (2\theta_1) M_1^5 \simeq 5.5. \times 10^{-22} \theta_1^2 \left(\frac{M_1}{{\rm keV}}\right)^5 \, {\rm s}^{-1}
\end{equation} 
\begin{figure}[h]
\begin{center}
\includegraphics[width=0.6\textwidth]{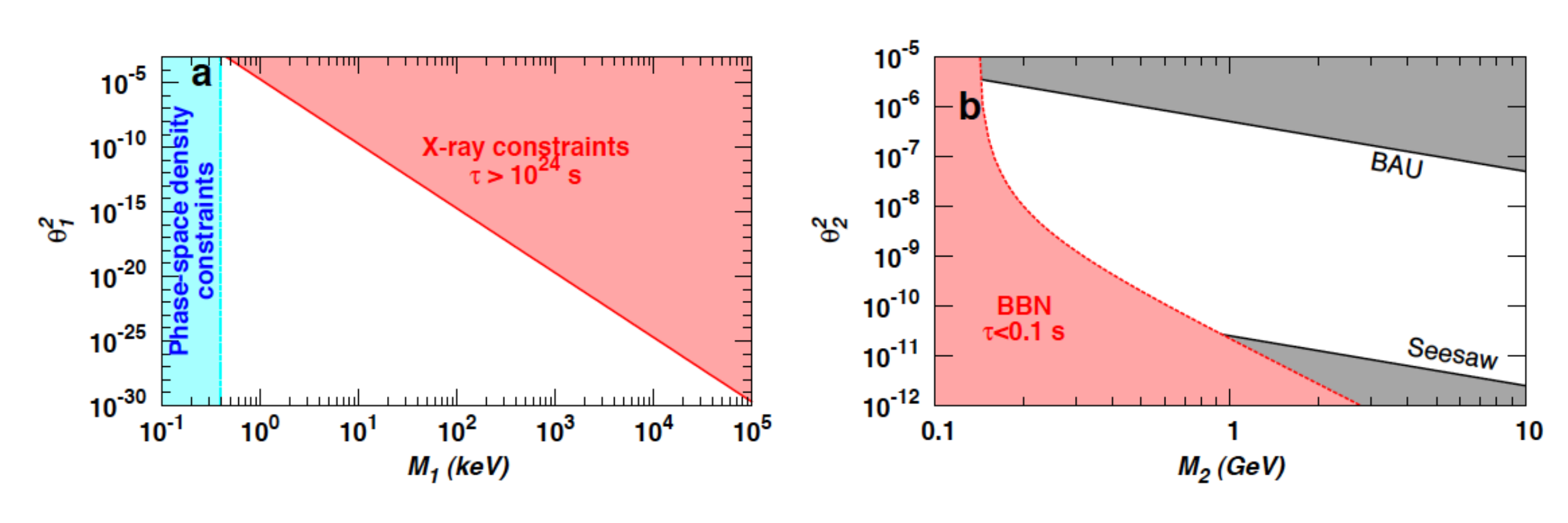}\hfill \includegraphics[width=0.4\textwidth]{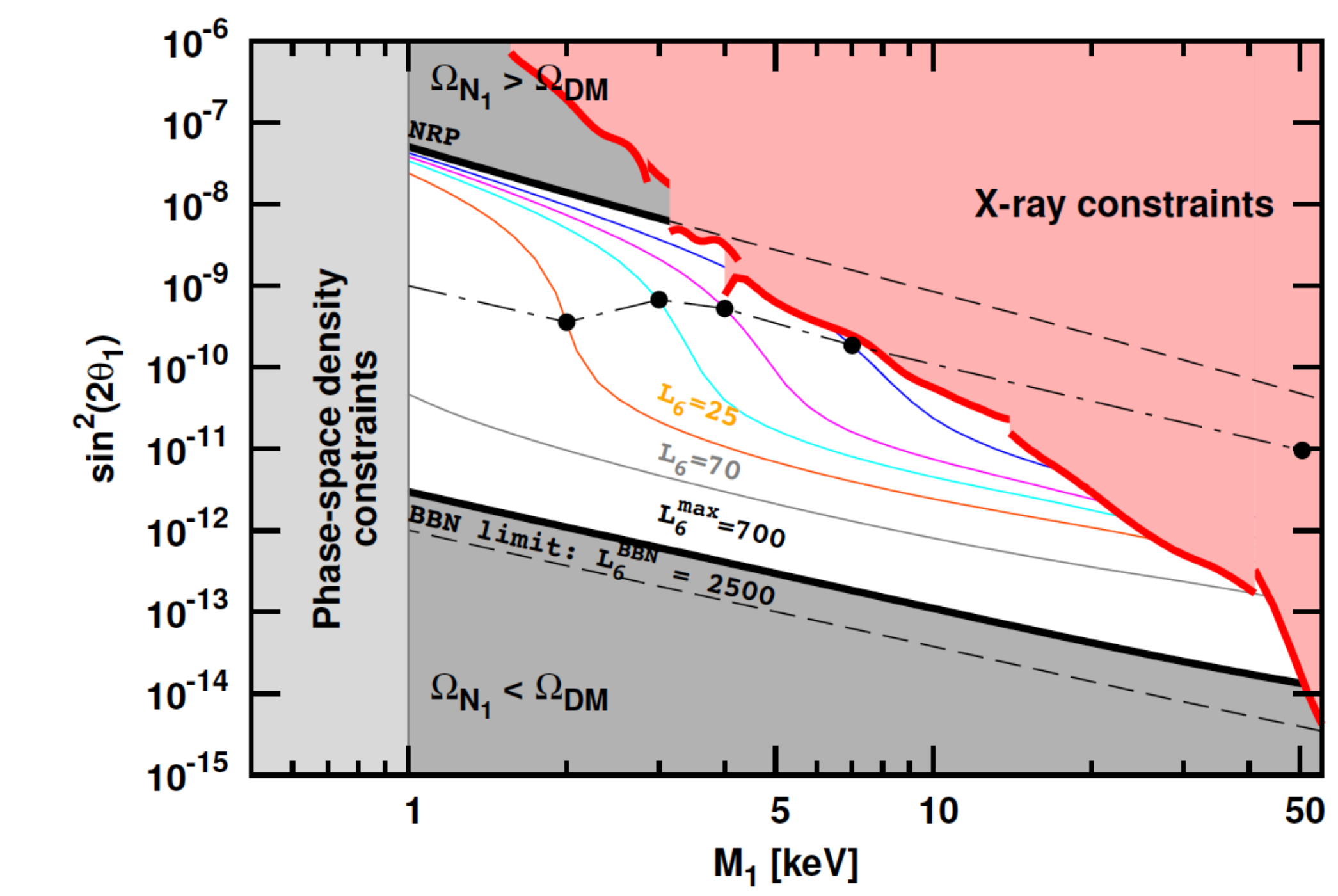}\end{center}
\caption{\label{fig:data} Astrophysical constraints for the $\nu$MSM model~\cite{nsm}.}
\end{figure}
which implies that, in order for the life time of $N_1$ to be larger than the Universe age, one must have $\theta_1^2  \le 1.8 \times  10^{-5} \left(\frac{M_1}{{\rm KeV}}\right)^{-5}$. The contributions to the mass matrix of the active neutrinos from this light sterile is estimated to be of order $\delta m_\nu \sim \theta_1^2 M_1 $, which can be within the experimental error for the solar mass differences of active neutrinos if $M_1 \ge 2~ {\rm keV}$.  The estimation of the production of the $N_1$ sterile neutrino in the Early universe is essential in order to have an idea of whether this candidate for DM satisfies the current astrophysical constraints~\cite{nsm,shapo}. Such an estimation requires taking into account the interactions of $N_1$ with the heavy degenerate $N_{2,3}$ neutrinos. Because the decaying $N_1$ produces a narrow spectral line
in the spectra of DM dominated astrophysical objects (e.g. galaxies), an important astrophysical constraint comes from the x-rays from such objects. 

The model is found consistent with all current astrophysical data, including x-ray constraints, Big Bang Nucleosynthesis (BBN) and Structure Formation data~\cite{nsm} (\emph{cf.} fig.~\ref{fig:data})..

\emph{Neutrino Condensates and dark Energy in the Universe:} Formation of fermion condensates dynamically in the Early Universe
as in Nambu-Jona-Lasinio model, has been considered by several authors. In this spirit one may consider models of (effective) four-fermion interactions of  sterile Majorana neutrino in the early Universe. The respective condensates may be formed, e.g. through a heavy scalar exchange. In the model of ref.~\cite{cond} it is argued that one light sterile neutrino (of mass of $O(10^{-3}$~eV)) may form the condensate at a late era of the early universe, and be responsible for the observed acceleration (and dark energy component of the energy budget) of the Universe. The model is argued to be consistent with the solar neutrino data.
Such ideas of neutrino forming condensates through self interactions and their relation to cosmology appear in other contexts, also related to baryogenesis, in a variety of works~\cite{worksde}.

\begin{figure}[h]
\begin{center}
\includegraphics[width=0.3\textwidth]{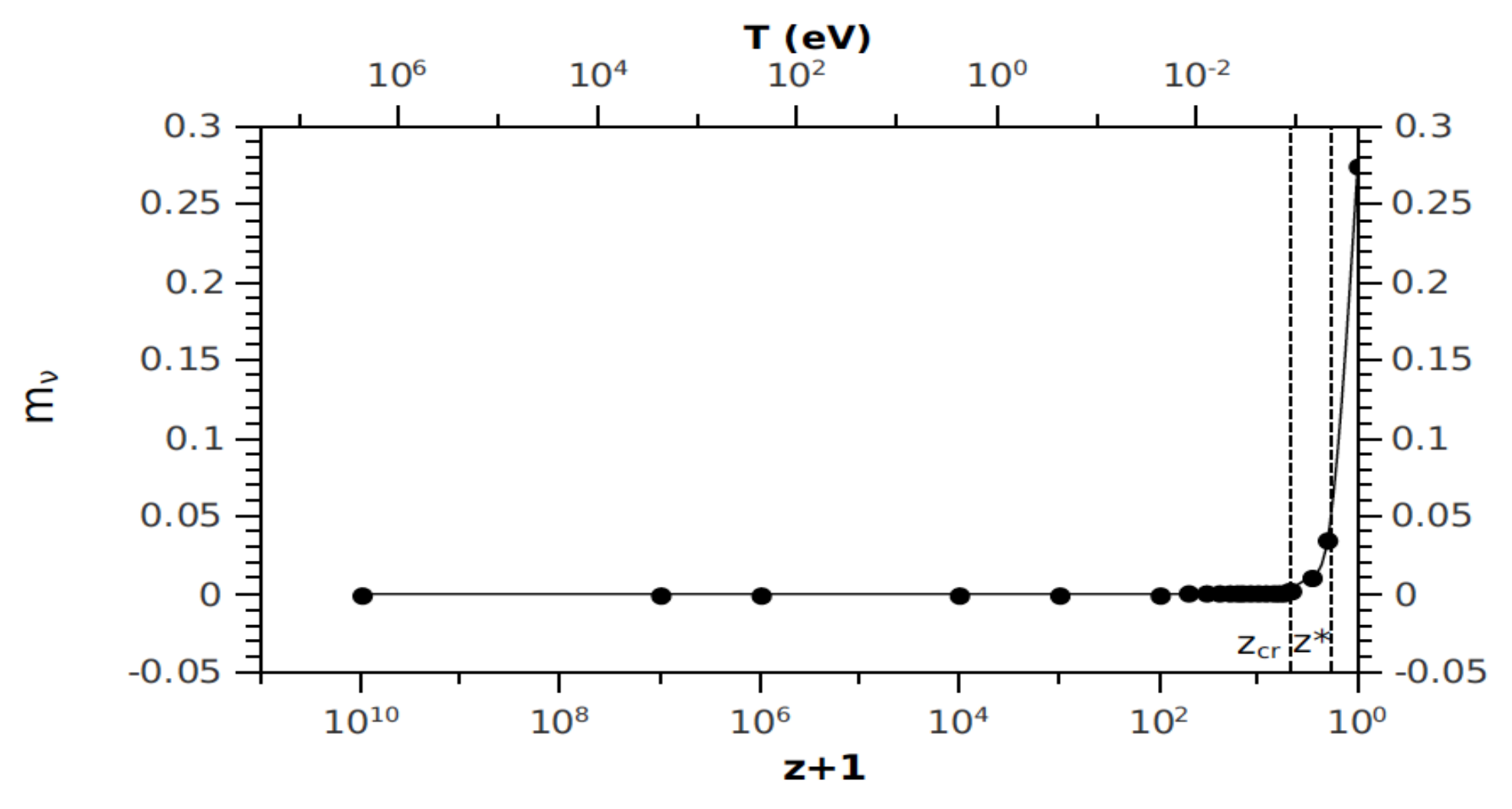}\hfill \includegraphics[width=0.4\textwidth]{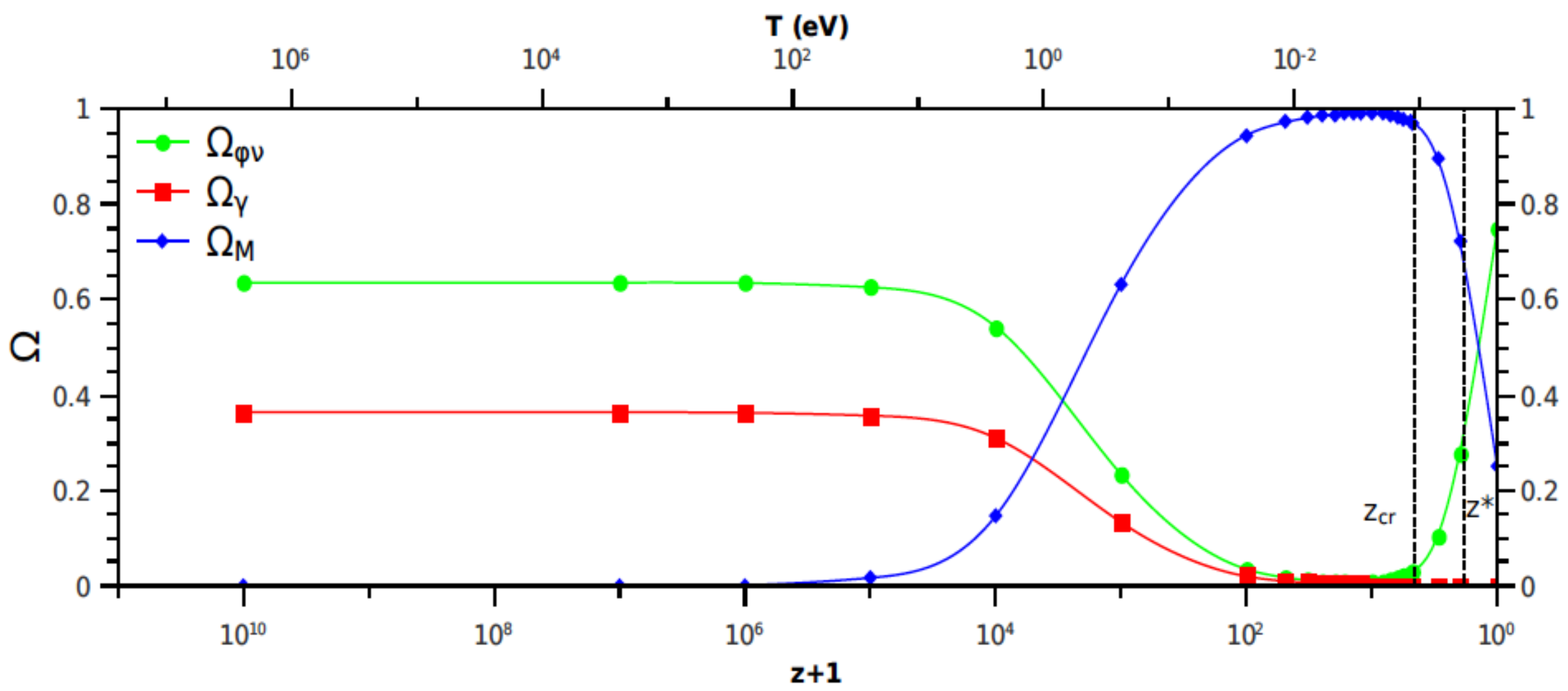}\hfill 
\includegraphics[width=0.3\textwidth]{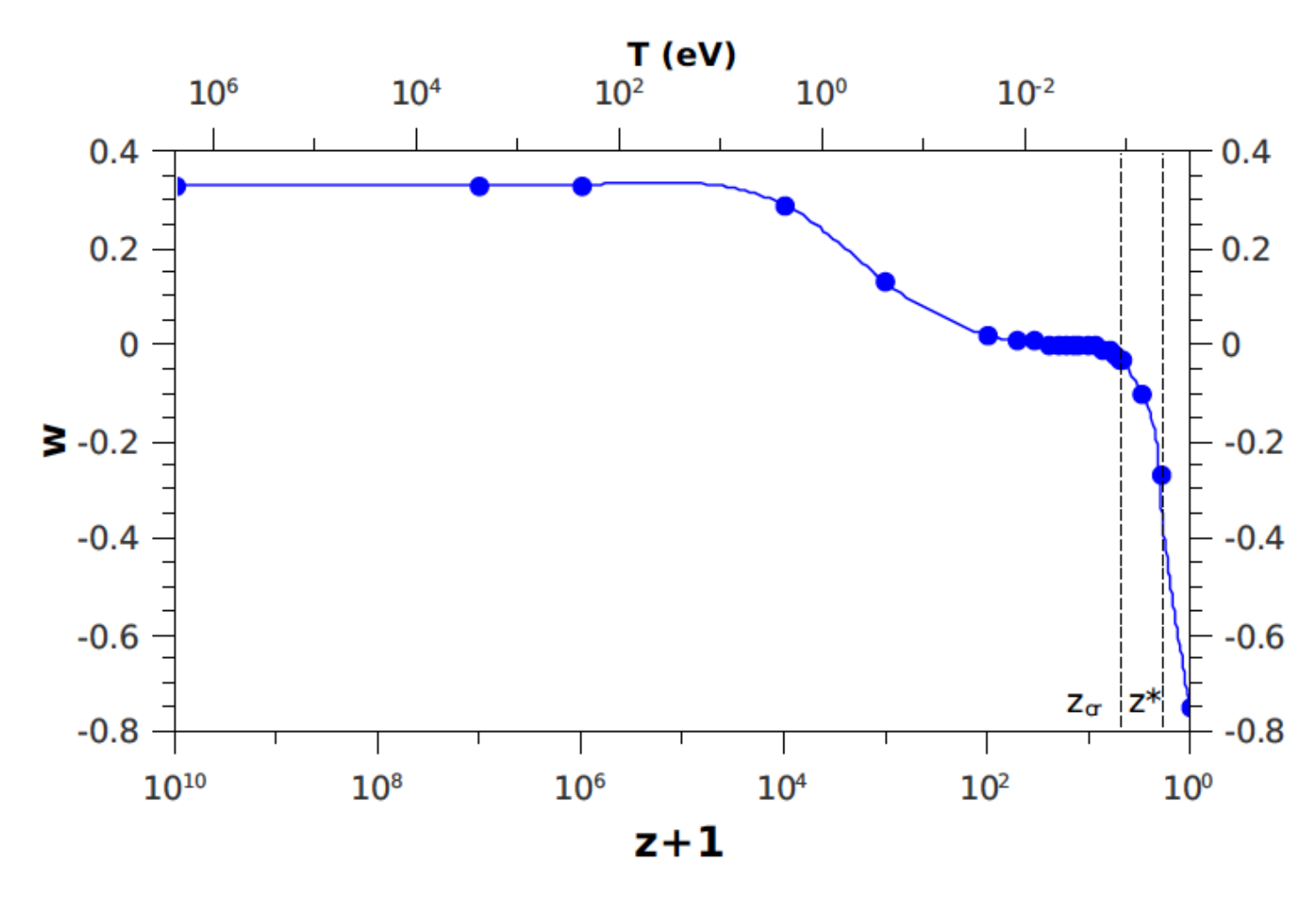}
\end{center}
\caption{\label{fig:mvn} \emph{Left:} The evolution of the mass of the neutrino vs T and the redshift $z$. \emph{Middle} Neutrino Dark Energy evolution vs that of Dark Matter . \emph{Right:} The relevant equation of state,
for a potential $U(\varphi) = M^{\alpha + 4}/\varphi^\alpha$, with $M = 2.39 \times 10^{-3}$ eV, $\alpha = 0.01$. From ref.~\cite{mvnc}.}
\end{figure}

\emph{Dark Energy from Mass Varying neutrinos:} This attractive idea~\cite{nelson} can be formulated simply as follows: one couple scalar cosmic fields with potential U($\varphi$, T), where T is the temperature of the Universe at a given era,  and massless fermions $\psi$ 
through Yukawa couplings  $g \int_0^{\beta= (k_BT)^{-1}}\, dt \int  a^3(t) d^3x\, \varphi \overline{\psi} \psi $. The (T-dependent) fermion mass $m=g \varphi_c (T)$ is acquired through minimization w.r.t. to $\varphi$ (i.e. $\varphi = \varphi_c (T)= <\varphi >$ at the minimum) of the effective potential density $ \Omega (\varphi) = U(\varphi) - \frac{1}{\beta a^3 V}{\rm log}Z_D(\varphi)$, where $Z_D$ is the partition function at temperature $T$. The neutrino mass (which can be significantly higher at early stages of the Universe evolution) decreases with the temperature T in an expanding Universe, and this has as a consequence the presence of a dark Energy fluid, whose equation of state resembles that of a cosmological constant at late eras~\cite{mvnc} (\emph{cf}. fig.~\ref{fig:mvn}). The models can be made consistent with current cosmologies for some choices of the potential $U(\varphi)$. However, a fundamental microscopic origin of such potentials is still lacking.

\section{CPT Violation in the Early Universe, active neutrinos and Baryon Asymmetry \label{sec:4}}

So far we have assumed that the CPT symmetry holds in the Early Universe, and this produces matter and antimatter in equal amounts.
An interesting idea is that during the Big Bang, one or more of the assumptions for the CPT theorem (Lorentz Invariance, unitarity and/or locality of interactions) breakdown (\emph{e.g}. due to quantum gravity influences that may be strong at such early times), which results in CPT Violations (CPTV) and a naturally induced matter-antimatter asymmetry, without the need for extra sources of CP violation, such as sterile neutrinos. The simplest possibilirty~\cite{dolgov} is through particle-antiparticle mass diffferences $m \ne \overline{m}$.  These would affect the (anti) particle distributions 
$f(E, \mu) = [{\rm exp}(E-\mu)/T) \pm 1]^{-1}$, $E^2 = p^2 + m^2$ and similarly for antimatter $m \rightarrow \overline{m}$, 
and thus generate a matter antimatter asymmetry in the relevant densities $n-{\overline n} = g_{d.o.f.} \int \frac{d^3 p}{(2\pi)^3}[f(E,\mu) - f(\overline{E},\mu)]$. Assuming~\cite{dolgov} quite reasonably that dominant contributions to Baryon asymmetry come from quarks-antiquarks, and that their masses are increasing, say, linearly with temperature $m \sim g T$, one estimates the induced baryon asymmetry by the fact that the maximum quark-antiquark 
mass difference is bounded by the current experimental bound on proton-antiproton mass difference, which is known to be less than $2 \cdot 10^{-9}$ GeV. This produces, unfortunately, too small BAU compared to the observed one. 

However, active neutrino-antineutrino mass differences alone may reproduce BAU; some phenomenological models in this direction have been considered in \cite{bbl}, considering for instance particle-antiparticle mass differences for active neutrinos  compatible with current oscillation data. 

But particle-antiparticle mass difference may not be the only way by which CPT is violated. As discussed in \cite{bm}, quantum gravity fluctuating effects that may be strong in the early unvierse, may act as an environment inducing decoherence for the (anti) neutrino , but with couplings between the particles and the environment that are different between the neutrino and antineutrino sectors. In \cite{bm} simple models of Lindblad decoherence
were considered, with zero decoherence parameters in the particle sector, and non trivial only in the antiparticle sector, and such that there was a mixed energy dependence (some of the coefficients (with dimension of energy) were proportional to the antineutrino energies, $\overline{\gamma}_i = (T/M_P)\, E$, while others were inversely proportional to them (and subdominant) $\overline{\gamma}_j = 10^{-24} \frac{1}{E} $, $j \ne i$). The model is phenomenological and its choice was originally motivated by fitting the LSND ``anomalous data'' in the antineutrino sector with the rest of the neutrino data. In this way one can derive an active (light) $\nu - \overline{\nu}$  asymmetry of order $\mathcal{A} = (n_\nu - n_{\overline{\nu}})/(n_\nu + n_{\overline{\nu}}) = \widehat{\gamma}_1 = \frac{T}{M_P} \cdot \frac{E}{\sqrt{\Delta m^2}}$. This Lepton number violation is communicated to the Baryon 
sector by means of B+L violating sphaleron
processes, as usual, and one can thus reproduce the observed BAU without the need for extra sources of CP violation and thus sterile neutrinos.
Unfortunately, at present such models lack microscopic understanding, but we think they are worth pursuing. For other scenarios of 
neutrino-anrineutrino CPT violation induced by local curvature effects in geometries of the early universe, and its connection to baryogenesis, see ref. \cite{mukho}.

\verb"Acknowledgments"
I thank the organisers of Nufact2011 for the invitation. 
This work was supported in part by the London Centre for
Terauniverse Studies (LCTS), using funding from the European Research
Council via the Advanced Investigator Grant 267352.

\end{document}